\documentclass{PoS}

\usepackage{lipsum}
\usepackage{lineno}

\usepackage{amsmath}
\usepackage{multicol}
\usepackage{mathtools}
\usepackage{cuted}
\usepackage{flushend}
\usepackage{amsmath}
\usepackage{graphicx}
\usepackage{bm}		
\usepackage{multirow}
\usepackage{enumitem}
\usepackage{graphicx}
\newcommand\varpm{\mathbin{\vcenter{\hbox{%
  \oalign{\hfil$\scriptstyle+$\hfil\cr
          \noalign{\kern-.3ex}
          $\scriptscriptstyle({-})$\cr}%
}}}}
\newcommand\varmp{\mathbin{\vcenter{\hbox{%
  \oalign{$\scriptstyle({+})$\cr
          \noalign{\kern-.3ex}
          \hfil$\scriptscriptstyle-$\hfil\cr}%
}}}}

\newcommand{\beq}{\begin{equation}}
\newcommand{\eeq}{\end{equation}}
\newcommand{\dif}[1]{\mathrm{d} #1 \,}

\renewcommand{\vec}[1]{\bm{#1}}
\definecolor{draftcolor}{rgb}{0.5,0.5,0.8}

\title{Measurement of azimuthal asymmetries in SIDIS \\ on unpolarized protons}

\ShortTitle{Measurement of azimuthal asymmetries in SIDIS on unpolarized protons}

\author{Andrea Moretti \\ 
        \rm{on behalf of the COMPASS Collaboration}\\
        \it{University of Trieste and INFN Trieste Section} \\
        E-mail: \email{andrea.moretti@ts.infn.it}}

\abstract{ The COMPASS Collaboration is measuring the asymmetries in the azimuthal distributions of positive and negative hadrons produced in Deep Inelastic Scattering (DIS) on unpolarized protons. The data have been collected in 2016 and 2017 with a 160 GeV/$c$ muon beam scattering off a liquid hydrogen target. The amplitudes of three modulations, $A_{UU}^{\cos\phi_h}$, $A_{UU}^{\cos 2\phi_h}$ and $A_{LU}^{\sin\phi_h}$ are measured as function of the Bjorken variable $x$, of the fraction of virtual photon energy carried by the hadron $z$, and of the hadron transverse momentum with respect to the virtual photon $p_{T}^{h}$. The relevance of azimuthal asymmetries lies in the possibility to get information on the intrinsic transverse momentum of the quark as well as on the still unknown Boer-Mulders parton distribution function. The preliminary results from 2016 data shown here confirm the strong kinematic dependencies observed in previous measurements conducted by COMPASS, HERMES and CLAS.}

\FullConference{23rd International Spin Physics Symposium - SPIN2018 -\\
		10-14 September, 2018\\
		Ferrara, Italy}

\begin{document}

\section{Introduction}

\begin{center}
    \begin{figure}[!h]
    \centering
    \includegraphics[width=0.5\textwidth]{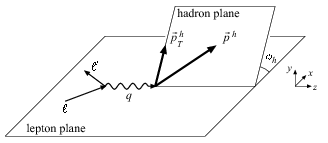}
    \caption{Definition of the gamma-nucleon system (GNS)}
            \label{fig:sketch}
\end{figure}

\end{center}

The DIS unpolarized differential cross section for the production of a hadron $h$, with transverse momentum $p_{T}^{h}$ in the gamma-nucleon system (GNS, Figure~\ref{fig:sketch}) is given, in the one-photon exchange approximation, by the expression \cite{Bacchetta:2006tn}:

\beq \label{eq:crosssect}
	\frac{\dif\sigma}{p_{T}^{h}\dif{p_{T}^{h}}\dif{x}\dif{y}\dif{z}\dif{\phi_h}}
    = \sigma_0\left( 1 + \epsilon_1 A_{UU}^{\cos\phi_h} \cos\phi_h
    	+ \epsilon_2 A_{UU}^{\cos2\phi_h} \cos2\phi_h
        + \lambda \epsilon_3 A_{LU}^{\sin\phi_h} \sin\phi_h \right)
\eeq
where $x$ is the Bjorken variable, $y$ is the fraction of beam energy carried by the virtual photon, $z$ is the fraction the photon energy carried by the hadron, $\phi_h$ is the hadron azimuthal angle measured from the lepton plane, $\lambda$ is the beam polarization and the kinematic factors $\epsilon$ are defined as:
\beq
	\epsilon_1 = \frac{2(2-y)\sqrt{(1-y)}}{1+(1-y)^2}, \hspace{1.5cm}
    \epsilon_2 = \frac{2(1-y)}{1+(1-y)^2}, \hspace{1.5cm}
    \epsilon_3 = \frac{2y\sqrt{1-y}}{1+(1-y)^2}.
\eeq

The amplitudes $A_{XU}^{f(\phi_h)}$, commonly referred to as azimuthal asymmetries, emerge as ratios of azimuthal-angle-dependent structure functions and the unpolarized part of the cross section. The subscripts $UU$ and $LU$ refer to unpolarized beam and target, and to longitudinally polarized beam and unpolarized target respectively. Particularly interesting are the $A_{UU}^{\cos\phi_h}$ and $A_{UU}^{\cos2\phi_h}$ asymmetries, whose expressions at leading twist in terms of transverse momentum dependent parton distribution functions (TMD-PDFs) and fragmentation functions (TMD-FFs) are given by:

\begin{equation}\label{eq:cos}
    A_{UU}^{\cos\phi_h} \propto \frac{2M}{Q}\mathcal{C} \left[ - \frac{(\hat{\bm{h}}\cdot \vec{p}_{\perp})\vec{k}_T^2}{M^2M_h} h_1^{\perp}H_1^{\perp} - \frac{(\hat{\bm{h}}\cdot \vec{k}_T)}{M} f_1D_1\right]
\end{equation}

\begin{equation}\label{eq:cos2}
    A_{UU}^{\cos2\phi_h} \propto \mathcal{C} \left[ - \frac{2(\hat{\bm{h}}\cdot \vec{k}_T)(\hat{\bm{h}}\cdot \vec{p}_{\perp})- \vec{k}_T\cdot \vec{p}_{\perp}}{MM_h} h_1^{\perp}H_1^{\perp}\right].
\end{equation}
In these expressions, $\hat{\bm{h}}$ is the normalized hadron transverse momentum vector with respect to the lepton plane in the GNS, $\vec{p}_{\perp}$ is the transverse momentum acquired by the hadron in the fragmentation process, $\vec{k}_T$ the quark intrinsic transverse momentum, $M$ the proton mass, $M_h$ the final hadron mass, $h_1^{\perp}$ the Boer-Mulders PDF, $H_1^{\perp}$ the Collins FF, $f_1$ and $D_1$ the unpolarized PDF and FF respectively. The symbol $\mathcal{C}[wfD]$ represents the convolution integral over the unobserved transverse  momenta $\vec{k}_T$ and $\vec{p}_{\perp}$ of the PDF $f$ and of the FF $D$, with a weight $w$.

The term proportional to $f_1$ in the $A_{UU}^{\cos\phi_h}$ asymmetry is the dominant one, and is related to the Cahn effect \cite{Cahn:1978se}, which is due to the non coplanarity of the scattering between lepton and proton in the initial state. Both $A_{UU}^{\cos\phi_h}$ and $A_{UU}^{\cos 2\phi_h}$ depend on the convolution of the Collins FF and the Boer-Mulders PDF. Together with information from $p_{T}^{h}$-dependent multiplicities \cite{Aghasyan:2017ctw}, unpolarized azimuthal asymmetries measured in SIDIS are thus interesting in view of a possible extraction of the Boer-Mulders PDF and evaluation of the intrinsic quark transverse momentum. Attempts of extracting them from published results, however, have not been conclusive so far \cite{Anselmino:2005nn,Barone:2005kt,Boglione_PhysRevD84}.
Extractions of azimuthal asymmetries have been performed in the past by COMPASS \cite{Adolph:2014pwc}, HERMES \cite{Airapetian:2012yg} and CLAS \cite{Osipenko:2008aa}. The COMPASS published results \cite{Adolph:2014pwc} have been obtained using data collected in 2004 on a $^6$LiD target with a positive muons beam. The amplitudes of the $\cos\phi_h$ and $\cos 2\phi_h$ modulations showed strong kinematic dependencies for both positive and negative hadrons. 

Recently, its has been realized by the COMPASS Collaboration \cite{Albi:2018fe} that hadrons from the decay of diffractively produced vector mesons exhibit azimuthal asymmetries which contribute to the measured asymmetries. A Monte Carlo simulation based on the Lund string model and on the $^3P_0$ mechanism simulating the Cahn effect has been successfully compared to the published COMPASS data on the $A_{UU}^{\cos\phi_h}$ asymmetries on deuterium.

Here we present preliminary results obtained from part of the data collected at COMPASS during the 2016 run with a 160 GeV/c muon beam (both $\mu^+$ and $\mu^-$ beams with balanced statistics, as required by the parallel Deeply Virtual Compton Scattering measurement) and a 2.5 m long liquid hydrogen target.

 \vspace{2cm} \ \\

\section{Extraction of azimuthal asymmetries}
\subsection{Data selection and available statistics}
 
\begin{figure}[!h]
    \includegraphics[width=0.32\textwidth]{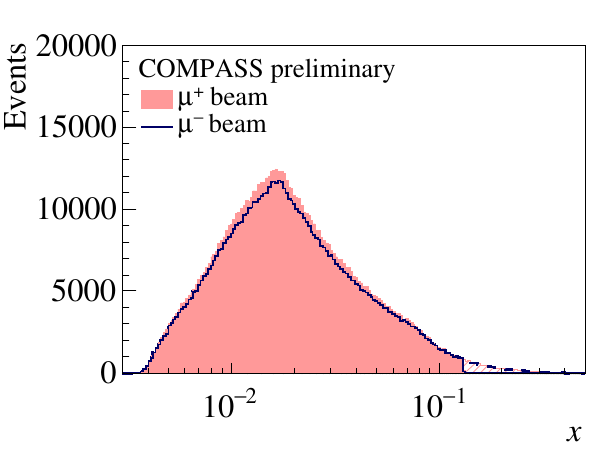} \includegraphics[width=0.32\textwidth]{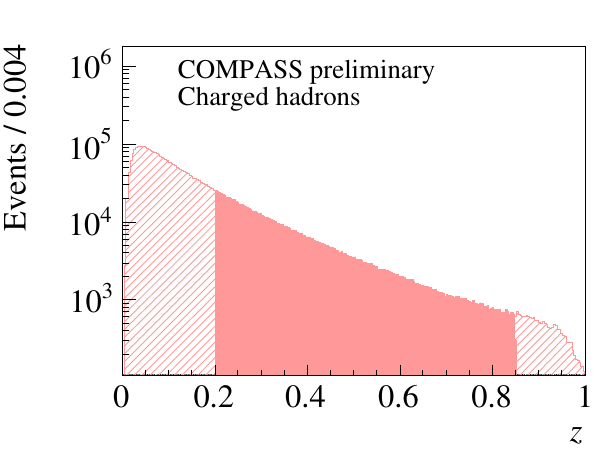}
    \includegraphics[width=0.32\textwidth]{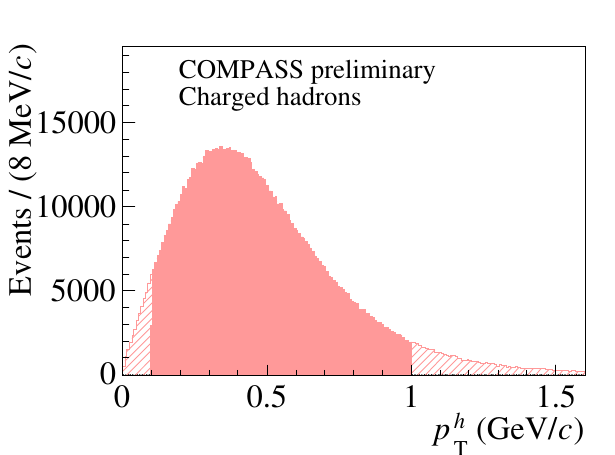}
    \caption{Kinematic distributions of the hadron sample. The (unnormalized) distribution of the Bjorken variable $x$ shows the remarkable balance between the samples collected with $\mu^+$ and $\mu^-$ beams. The $z$ and $p_{T}^{h}$ distributions are also given. The shaded regions are removed with proper cuts. }
    \label{fig:kin_dist}
\end{figure}

\begin{figure}[!h]
    \centering
    \includegraphics[width=0.55\textwidth]{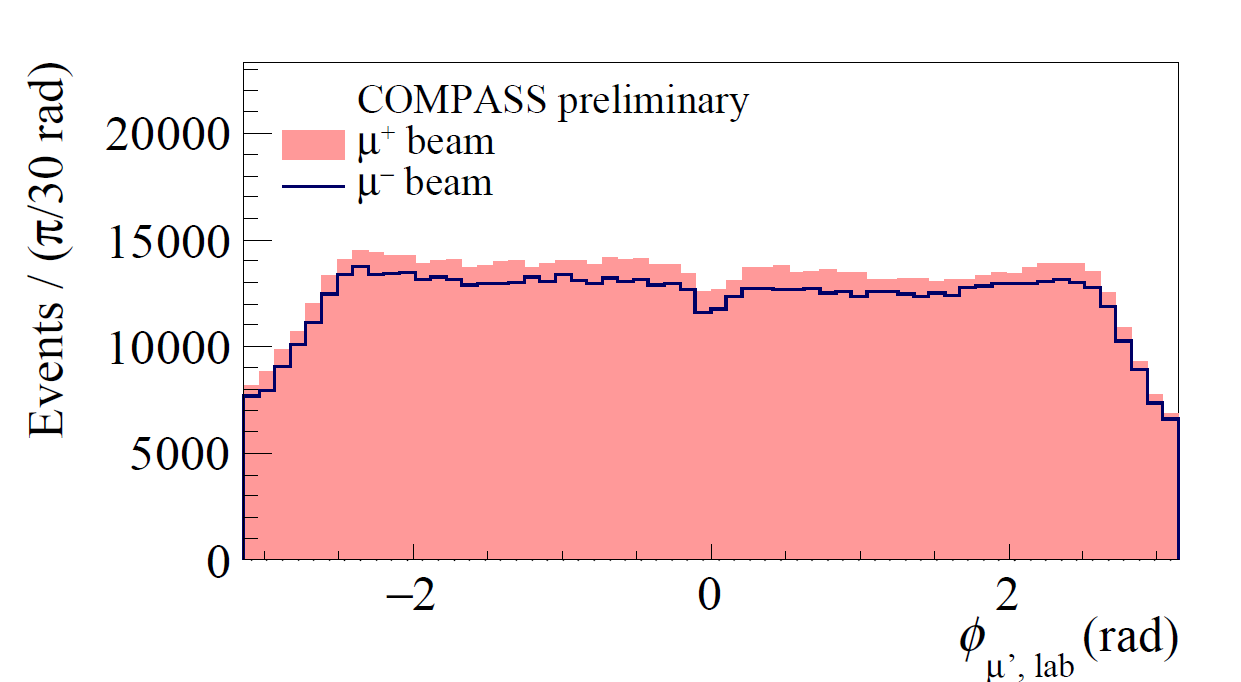} 
    \caption{Distribution of the scattered muon azimuthal angle $\phi_{\mu^\prime lab}$.The (unnormalized) distribution of the Bjorken variable $x$ shows the remarkable balance between the samples collected with $\mu^+$ and $\mu^-$ beams.}
    \label{fig:phi_muprim}
\end{figure}

The selection of DIS events has been performed asking for a squared photon virtuality $Q^2>1$ (GeV/$c)^2$, $0.2<y<0.9$, mass of the hadronic final state $W>5$ GeV/$c^2$ and $0.003<x<0.13$. To reduce the acceptance corrections, a limit was set on the polar angle in the laboratory frame of the virtual photon: $\theta_{\gamma}^{lab}<60$ mrad. Hadrons have been selected by requiring the material integrated over the path of the reconstructed trajectories not to exceed 10 radiation lengths $X_0$. Kinematic cuts on $z$ ($0.2<z<0.85$, to select the current fragmentation region and reduce exclusive events) and on $p_{T}^{h}$ (0.1 GeV/$c<p_{T}^{h}<$1.0 GeV/$c$, for good $\phi_h$ definition and to avoid larger acceptance corrections) have also been applied. The acceptance in the azimuthal angle of the scattered muon is almost flat (Figure \ref{fig:phi_muprim}). Plots of kinematic distributions as functions of $x$, $z$ and $p_{T}^{h}$ are given in Figure \ref{fig:kin_dist}. The final sample consists of almost 270 000 positive and 215 000 negative hadrons with a positive muon beam and almost 250 000 positive and 200 000 negative hadrons for negative muon beam. 

\vspace{0.5 cm} \ \\

\vspace{1.5cm} \ \\

\subsection{Procedure}

The unpolarized azimuthal asymmetries have been extracted alternatively as a function of $x$, $z$ and $p_{T}^{h}$. For each kinematic bin, the distribution of the azimuthal angle of the hadron in the GNS $\phi_h$ has been corrected for acceptance, obtained with a Monte Carlo simulation based on LEPTO generator, and defined as the ratio of reconstructed and generated hadrons:

\begin{equation}
    Acc(v,\phi_h) = \frac{n_{rec}^{h}(v,\phi_h)}{n_{gen}^{h}(v,\phi_h)}
\end{equation}
where $v$ is either $x$, $z$ or $p_{T}^{h}$. The acceptance corrections are smaller than 10\% in every kinematic bin. The central region of the $\phi_h$ distributions ($-\pi/8 < \phi_h < \pi/8$) has been removed to get rid of the contribution of radiative events. Then, the acceptance-corrected distributions have been fitted to extract the azimuthal asymmetries.

\clearpage
\newpage

\subsection{Results}

\begin{figure}[!h]
    \centering
    \includegraphics[width=0.65\textwidth]{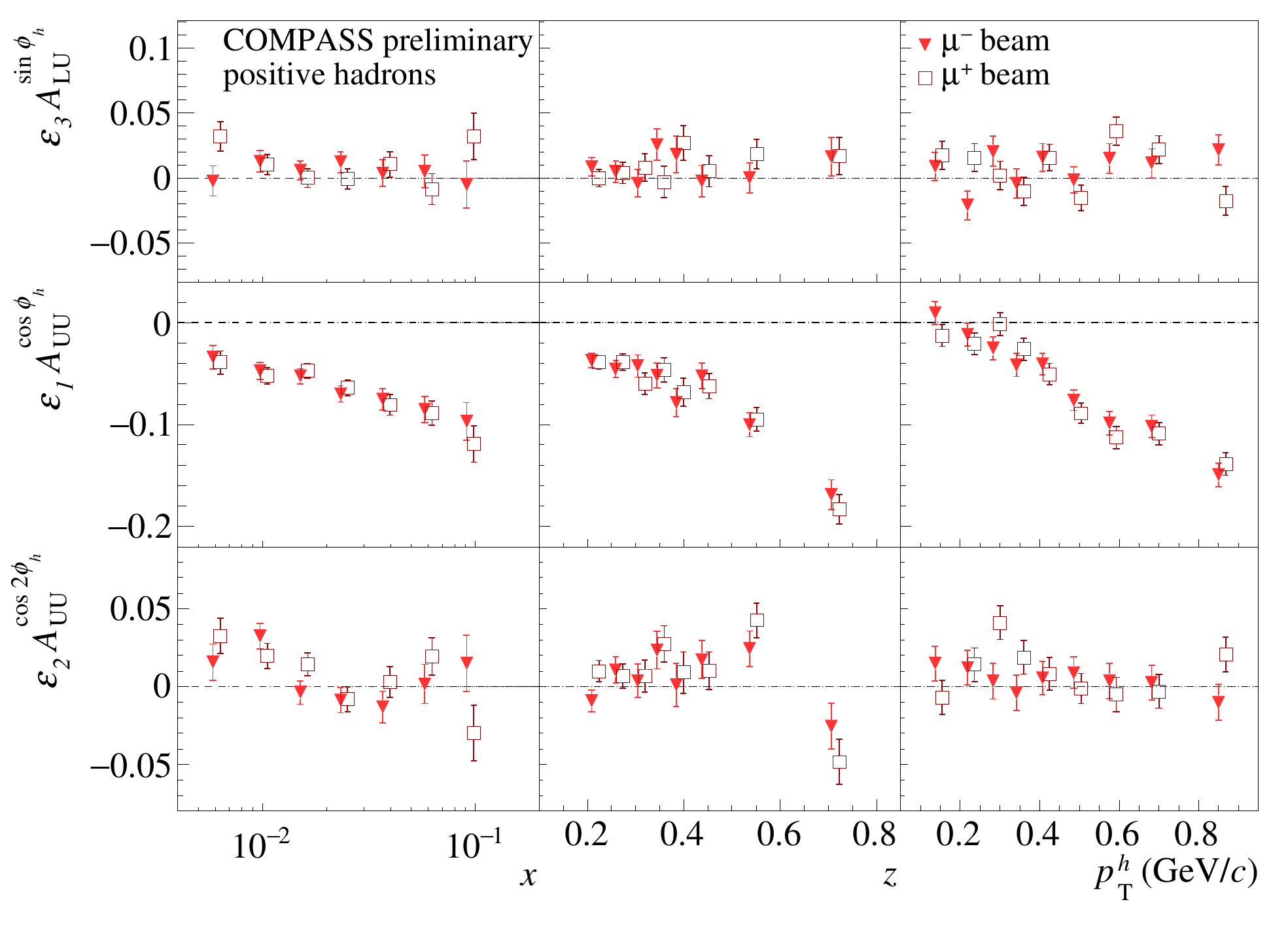}
    \caption{Azimuthal asymmetries for positive hadrons for the samples of data collected with positive and negative muon beam charges. Uncertainties are statistical only.}
    \label{fig:azyplus}
\vspace{0.5cm}  
    \centering
    \includegraphics[width=0.65\textwidth]{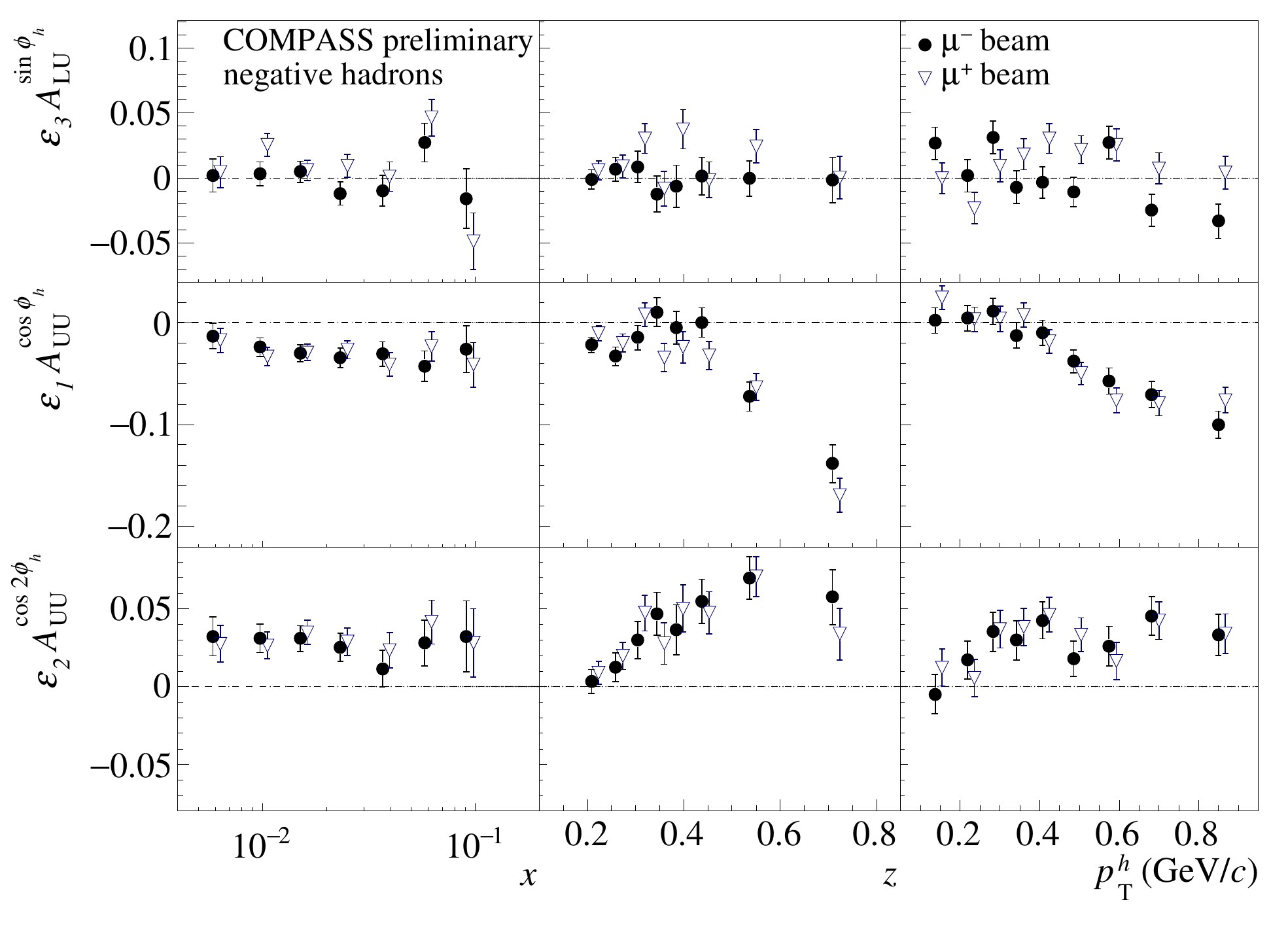}
        \caption{Azimuthal asymmetries for negative hadrons for the samples of data collected with positive and negative muon beam charges. Uncertainties are statistical only.}
    \label{fig:azymin}
\end{figure}

\begin{figure}[!h]
    \centering
    \includegraphics[width=0.78\textwidth]{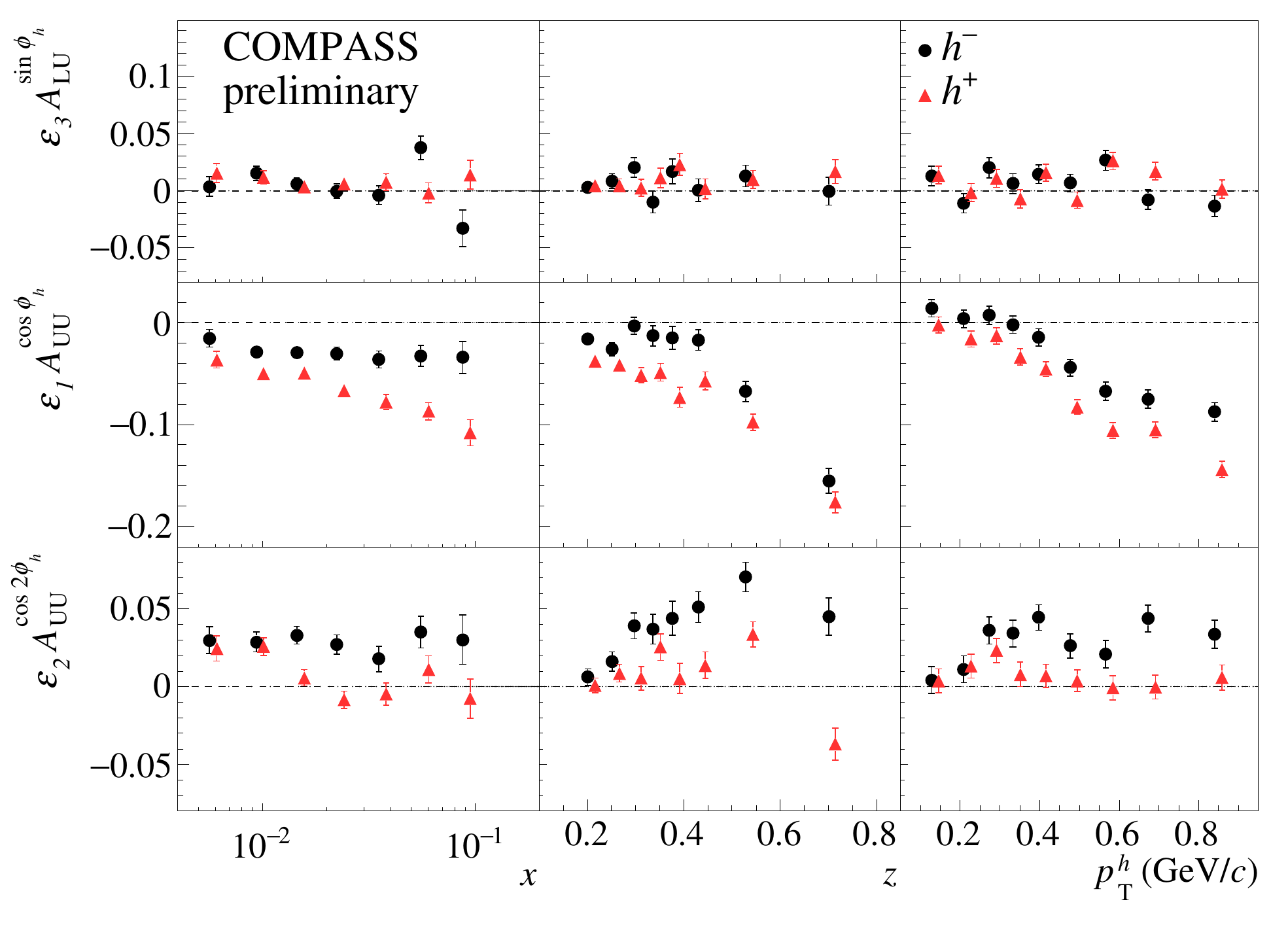}
    \caption{Azimuthal asymmetries for positive and negative hadrons, obtained after merging the results from the positive and negative muon beam charges samples. Uncertainties are statistical only.}
    \label{fig:azytot}
\end{figure}

The azimuthal asymmetries, multiplied by the corresponding kinematic factors $\epsilon$, are shown for positive and negative muon beams in Figures \ref{fig:azyplus} and \ref{fig:azymin} for positive and negative hadrons respectively. As expected, the asymmetries obtained with positive and negative muon beams are compatible, and consequently they have been merged. The obtained asymmetries are given in Figure \ref{fig:azytot}. While $\varepsilon_3 A_{LU}^{\sin\phi_h}$ shows a flat trend in all the three kinematic variables, $\varepsilon_1 A_{UU}^{\cos\phi_h}$ and $\varepsilon_2 A_{UU}^{\cos 2\phi_h}$ show strong kinematic dependencies. This is particularly the case for the $\cos\phi_h$ asymmetry, which reaches absolute values of almost 20\% when looked at as a function of $z$. The strong kinematic dependencies, already observed in the past, are therefore confirmed in the current analysis. 

\vspace{0.5cm}\ \\

\begin{figure}[!h]
    \centering
    \includegraphics[width=0.45\textwidth]{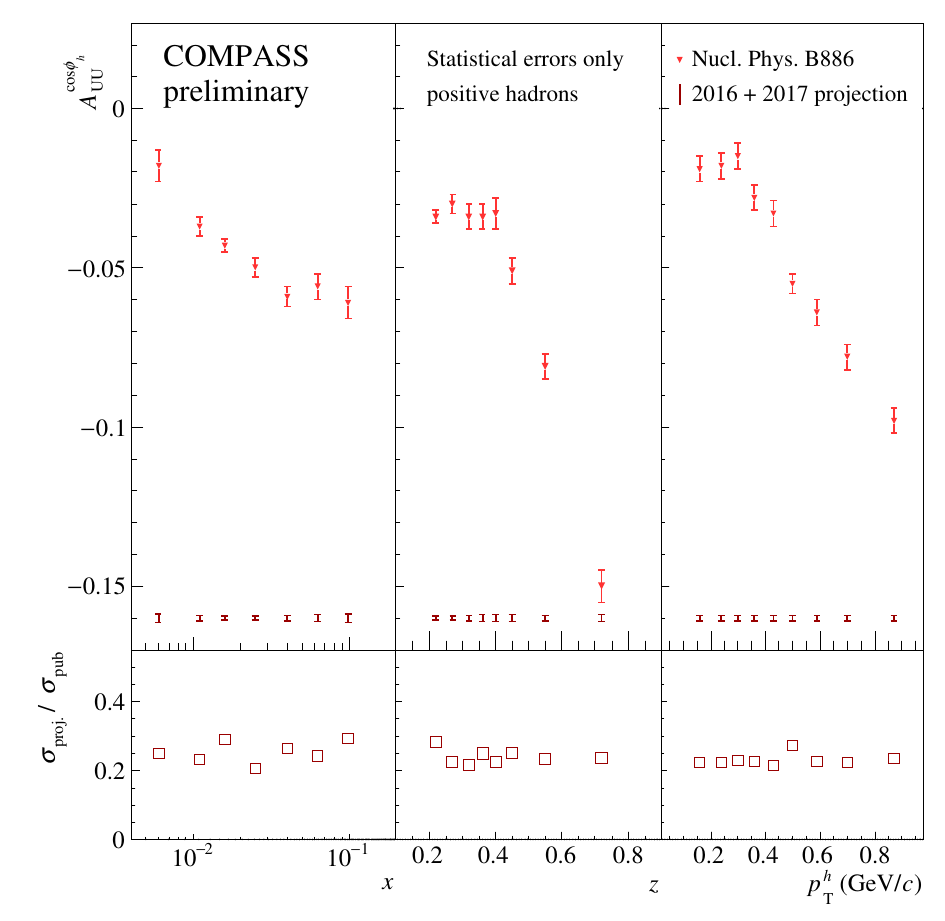} \includegraphics[width=0.45\textwidth]{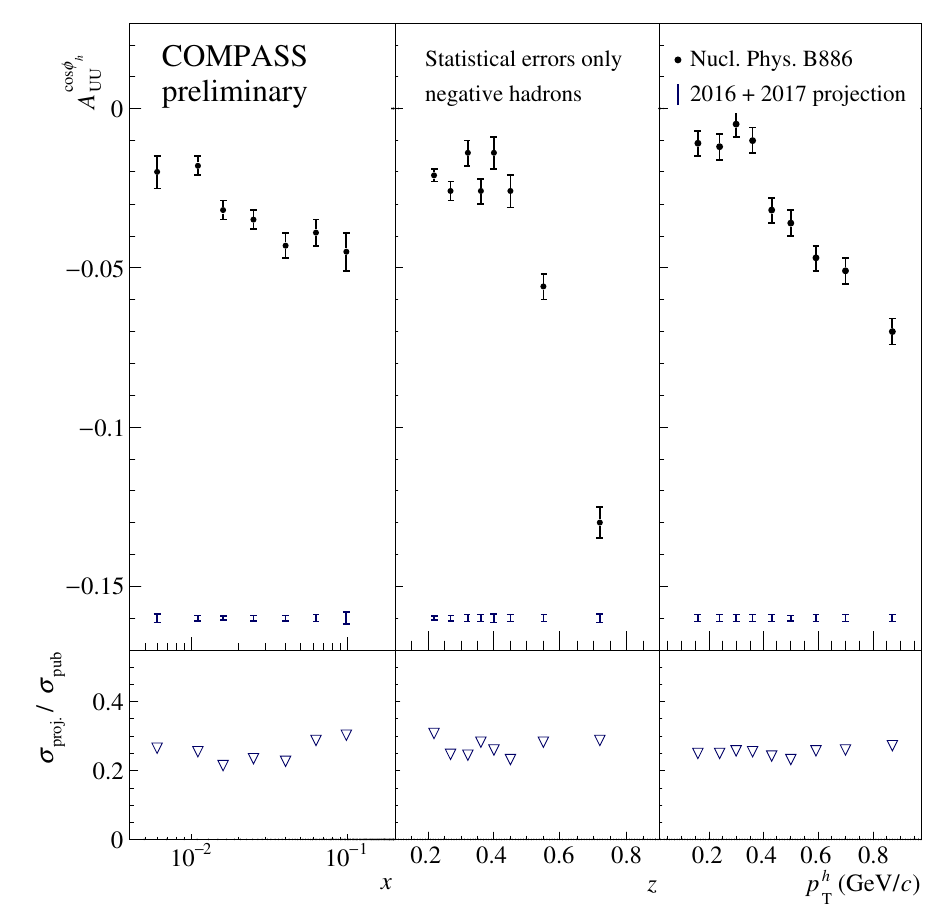}
    \caption{Comparison of $\cos\phi_h$ asymmetry projected uncertainty for the whole 2016+2017 data with the published one \cite{Adolph:2014pwc}, for positive and negative hadrons.}
    \label{fig:err_proj}
\end{figure}

The subsample used for this preliminary analysis is limited, corresponding to only about 4\% of the full data set collected in 2016 and 2017. In Figure \ref{fig:err_proj} one can see the projected uncertainty for the full sample as compared to the published results. Assuming a stable performance of the experimental apparatus, an overall gain by a factor of five is expected. Also, the characteristics of the spectrometer used for the 2016 and 2017 measurements are such that systematic uncertainties will also be largely reduced.

\newpage
\clearpage

\section{Conclusions and perspectives}

We presented COMPASS preliminary results on unpolarized azimuthal asymmetries for positive and negative hadrons produced with a positive and negative muon beam on an unpolarized proton target. The asymmetries have been extracted as function of Bjorken variable $x$, fraction of virtual photon energy carried by the hadron $z$ and hadron transverse momentum with respect to the virtual photon $p_{T}^{h}$. The one-dimensional analysis has been performed on part of the data collected at COMPASS in 2016, confirming the strong kinematic dependencies of azimuthal asymmetries already observed at COMPASS with a deuteron target, at HERMES and at CLAS. The current analysis will be extended to the full 2016 and 2017 data set, allowing for a reduction of the statistical uncertainties by a factor five with respect to the current results. Systematic uncertainties will also be reduced. The new data will also allow a deep investigation of the contribution to the azimuthal asymmetries from hadron produced in the decay of diffractive vector mesons. Together with information from other measurements, the azimuthal asymmetries may lead to an extraction of the still unknown Boer-Mulders TMD and of the quark intrinsic quark momentum $k_T$. 
\bigskip

\bibliographystyle{jhep.bst}

\providecommand{\href}[2]{#2}\begingroup\raggedright\endgroup

\end{document}